\documentstyle[11pt,mrs2001,epsfig]{article}
\begin{document}
\title{THE COSMIC SHEAR STIS PARALLEL PROGRAM - FIRST RESULTS}

\author{H. H\"{A}MMERLE$^{1,2}$, J.-M. MIRALLES$^{1,3}$,
P. SCHNEIDER$^{1,2}$, T. ERBEN$^{2,4,5}$, R.A.E.~FOSBURY$^{3}$,
W. FREUDLING$^{3}$, N. PIRZKAL$^{3}$,
S.D.M. WHITE$^{2}$} 
\affil{$^1$Institut f{\"u}r Astrophysik und Extraterrestrische
 Forschung der Universit{\"a}t Bonn, Auf dem H{\"u}gel 71, D-53121
 Bonn, Germany}
\affil{$^2$Max-Planck-Institut f{\"u}r Astrophysik, Karl-Schwarzschild
 Str. 1, D-85748 Garching, Germany}  
\affil{$^3$ST-ECF, Karl-Schwarzschild Str. 2,
 D-85748 Garching, Germany}  
\affil{$^4$Institut d'Astrophysique de Paris, 98bis Boulevard Arago,
 F-75014 Paris, France }  
\affil{$^5$Observatoire de Paris, DEMIRM 61, Avenue de l'Observatoire,
 F-75014 Paris, France}

\begin{abstract}
Since the Universe is inhomogeneous on scales well below the Hubble
radius, light bundles from distant galaxies are deflected and distorted
by the tidal gravitational field of the large-scale matter
distribution as they propagate through the Universe. Two-point
statistical measures of the observed ellipticities, like the
dispersion within a finite aperture or the ellipticity
cross-correlation, can be related to the power spectrum of the
large-scale structure. The measurement of cosmic shear (especially on
small angular scales) can thus be used to constrain cosmological
parameters and to test cosmological structure formation in the
non-linear regime, without any assumptions about the relation between
luminous and dark matter.
In this paper we will present preliminary cosmic shear measurements on
sub-arcminute scales, obtained from archival STIS parallel data. The
high angular resolution of HST, together with the sensitivity and
PSF-stability of STIS, allows us to measure cosmic shear along many
independent lines-of-sight.  Ongoing STIS parallel observations,
currently being carried out in the frame of a big GO program (8562+9248),
will greatly increase the number of available useful fields and will
enable us to measure cosmic shear with higher accuracy on sub-arcminute
scales.
\end{abstract}

\def\ave#1{\left\langle #1 \right\rangle}
\def\Pg{P^\gamma}
\def\cs{\ave{\overline{\gamma}^2} }
\def\csn{\overline{\gamma}_n^2 }
\def\sigcs{\sigma_{\ave{\overline{\gamma}^2}} }

\section{STIS Parallel Data}
Since June 1997, parallel observations using the Space Telescope Imaging
Spectrograph (STIS) on the Hubble Space Telescope were  taken. The data were
non-proprietary and were made available immediately. 

The STIS CCD is sensitive to wavelengths ranging from 2500 to 11000 \AA\
and its field of view is $51^{\prime\prime} \times
51^{\prime\prime}$. The STIS CCD pixel size is $50\mathrm{mas}$. 

For the cosmic shear project we selected data sets from June 1997 to
1998 which were taken in
the CLEAR filter (unfiltered) and  in  CR--SPLIT mode. They had to be
unbinned, and the associated ``jitter ball'', which is a measure of
how well the telescope was tracking, was required to have an rms value smaller
than $5\mathrm{mas}$. Images which were taken consecutively during a
single telescope visit and which were offset by no more than one
quarter of the field of view were grouped into what we refer to as STIS
association and coadded using ``Drizzle'' (see Fruchter \& Hook
1998). We obtained  498 coadded associations, which are available at
\texttt{http://www.stecf.org/projects/shear/}.  
A detailed description of the data reduction and coaddition procedure
can be found in \cite{PapI}.

Starting from these associations, 51 stellar fields and 121 independent
galaxy fields were identified.  
The fields were analysed using both SExtractor \cite{BA96} and
a modified version of the IMCAT package (see \cite{KSB} (hereafter KSB),
\cite{Eea01}). Size and shape parameters were taken from IMCAT,
since it was designed specifically to measure robust ellipticities and
allows for the correction of measured image ellipticities for shape
distortion introduced by the PSF. Positions and magnitudes are estimated
with SExtractor. The mean number of selected galaxies per association
is 18 on the galaxy fields.  

\section{PSF correction}
The shape of the PSF anisotropy was estimated on the stellar fields
using the KSB complex 
ellipticity parameter $e$, which is calculated from weighted second
order brightness moments. 

The total response of a galaxy ellipticity to a shear and the PSF is
given by
\begin{equation} 
 e -e_{\mathrm{s}} = \Pg \gamma + P_{\mathrm{sm}} q^*,
\end{equation}
where $e$ and $e_{\mathrm{s}}$ are the observed and intrinsic
ellipticities, respectively, and 
$ \Pg = P_{\mathrm{sh}} - \left(P_{\mathrm{sh}}/P_{\mathrm{sm}}
 \right)^* P_{\mathrm{sm}}$,
where the shear and the smear polarizability tensors $P_{\mathrm{sh}}$
and $P_{\mathrm{sm}}$ can be calculated from the galaxy light
profile (see KSB). The stellar anisotropy kernel $q^*$ which is needed
to correct for PSF anisotropy can be calculated by noting that for
stars $e_{\mathrm{s}}^*=0$ and $\gamma^*=0$, so that  
\begin{equation} 
 q^* = (P_{\mathrm{sm}}^*)^{-1} e^* ,
\end{equation}

\begin{figure}
\plotone{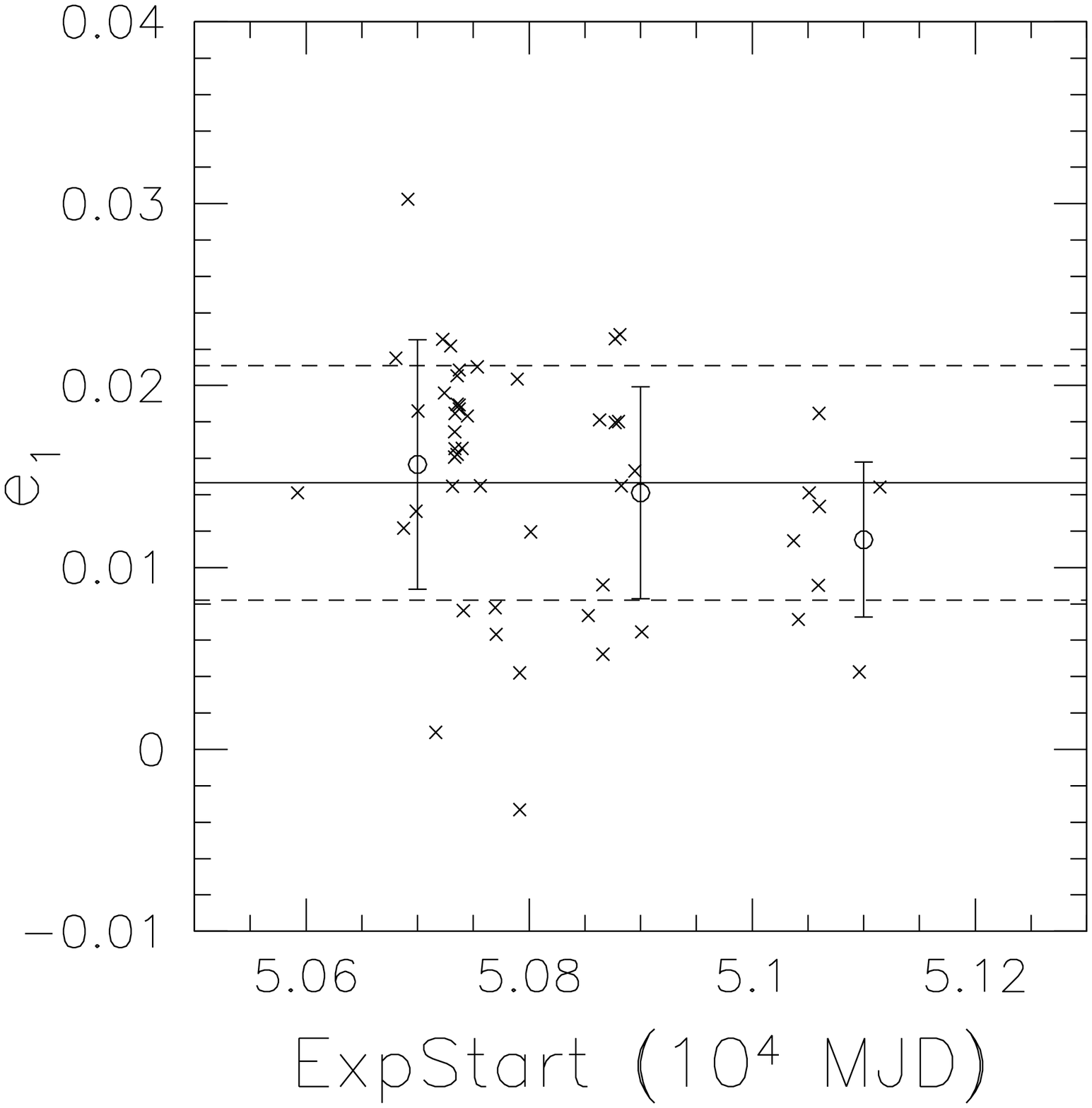}{7cm}\hspace{1cm}
\plotone{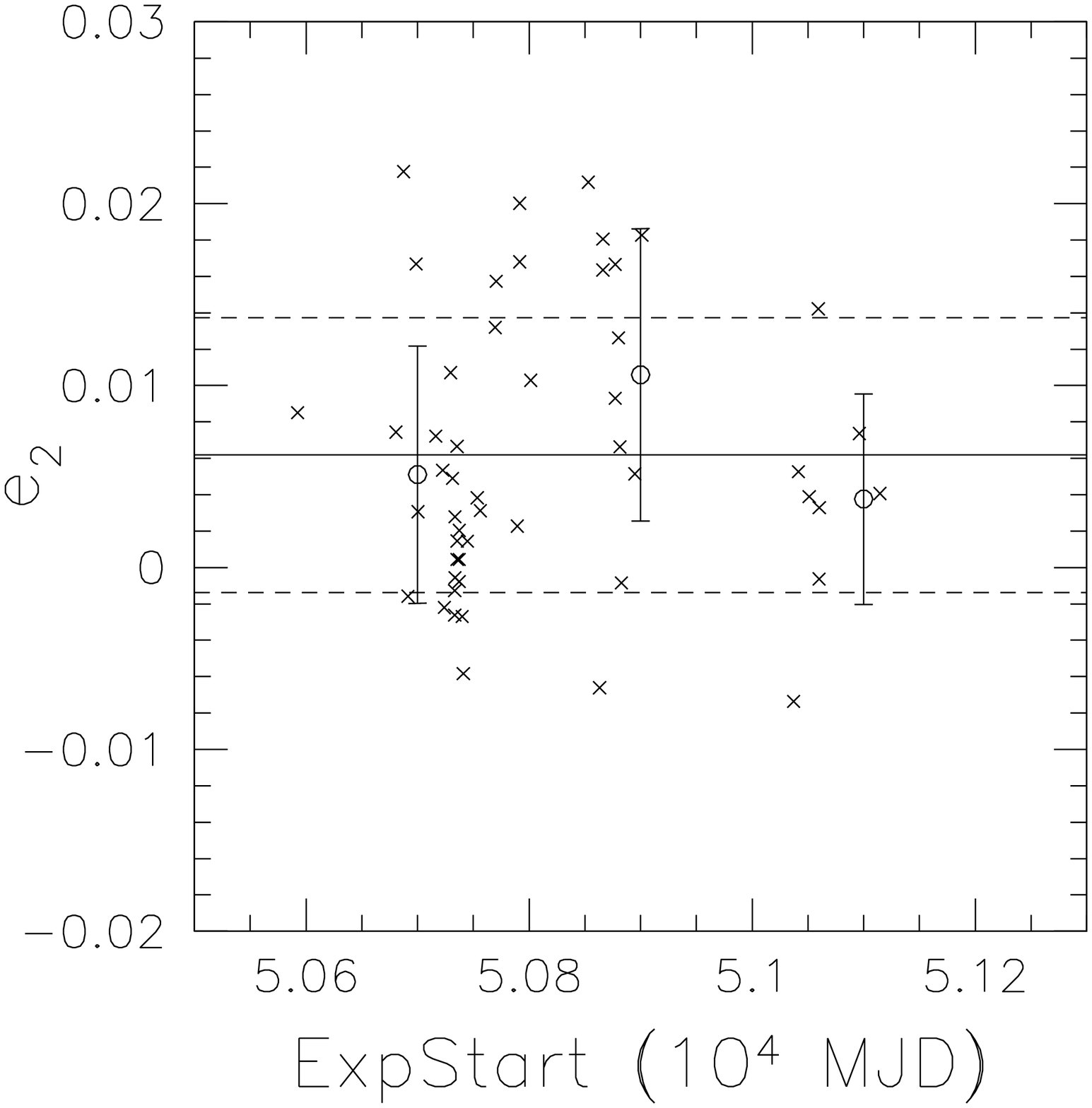}{7cm}
\caption{Mean of the ellipticity components $e_1$ (left) and $e_2$
(right) of the star fields vs. exposure start time (MJD). The straight
lines show the mean over all the fields, the dashed lines show the
$1\sigma$ dispersion. The circles show the mean over stars in bins
between $5.06\times 10^4$,   $5.08\times 10^4$, $5.10\times 10^4$,
$5.12\times 10^4$ (MJD) with $1\sigma$ error bars. }
\end{figure}

In Fig.~1 the mean values over the whole field of the two ellipticity
components of stars are shown as a function of
the exposure date. The mean ellipticity of all fields is $\approx1\%$,
which is suffiently small to not significantly affect our cosmic shear
analysis. If we divide the 
star fields into time intervals, the mean ellipticities agree with each
other at the $1\sigma$ level, therefore the anisotropy can be
considered to be constant over the time period covered. 

In addition to the variation from field to field (i.e. in time), we
also find a spatial  variation of the PSF within individual
fields. This effect is shown in Fig.~2, middle left panel. We fit the
ellipticities with a second-order polynomial at the 
position $\vec{\theta}=(x,y)$ on the CCD:

\begin{equation}
 e_\alpha(\vec{\theta}) = a_{\alpha0} + a_{\alpha1}x + a_{\alpha2}x^2
           + a_{\alpha3}y + a_{\alpha4}y^2 + a_{\alpha5}xy,
\end{equation}
where $\alpha=1,2$.  

\begin{figure}[!t]
\begin{minipage}{10cm}
 \plotone{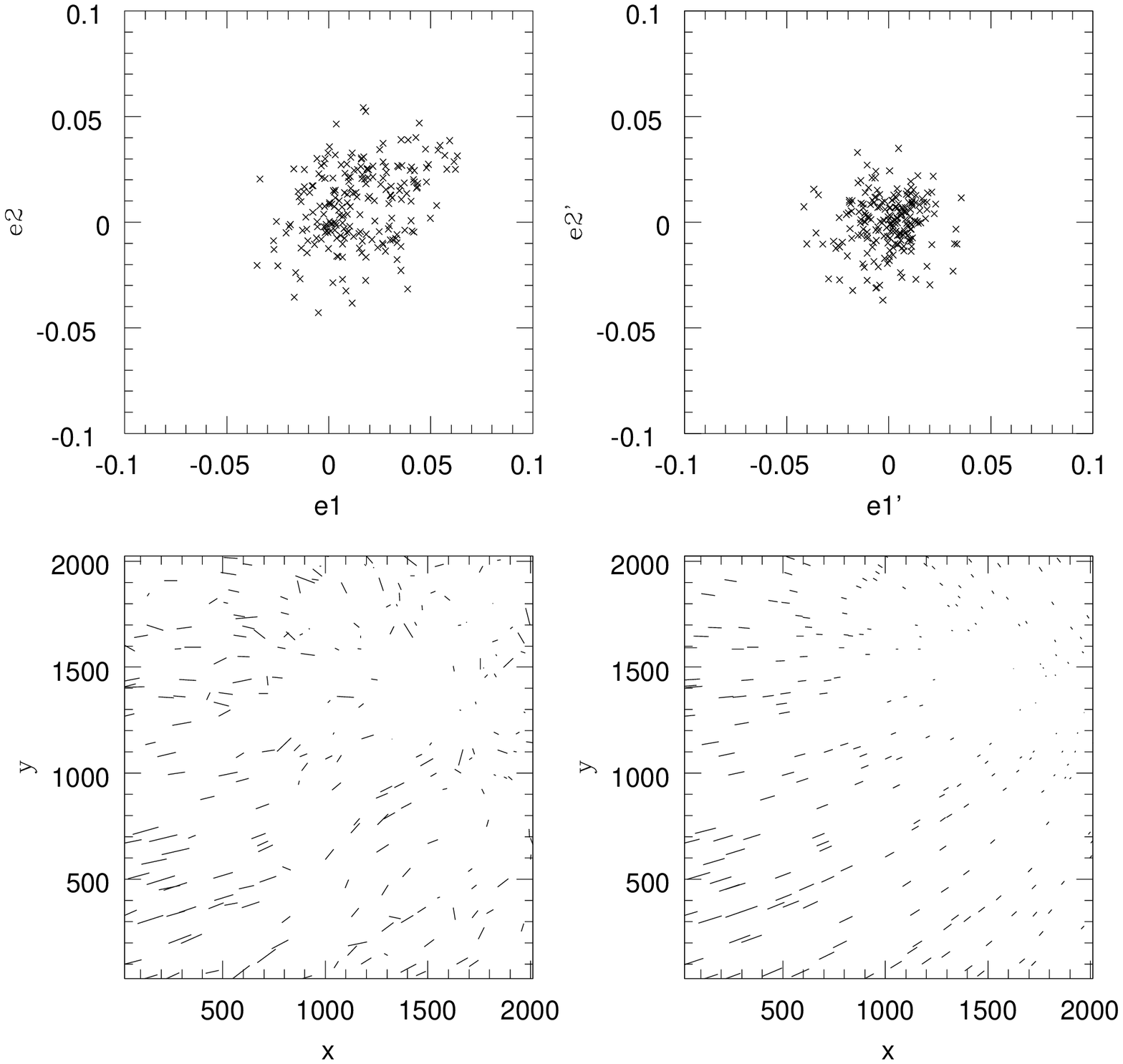}{10cm}\\[-0.3cm]
 \plotone{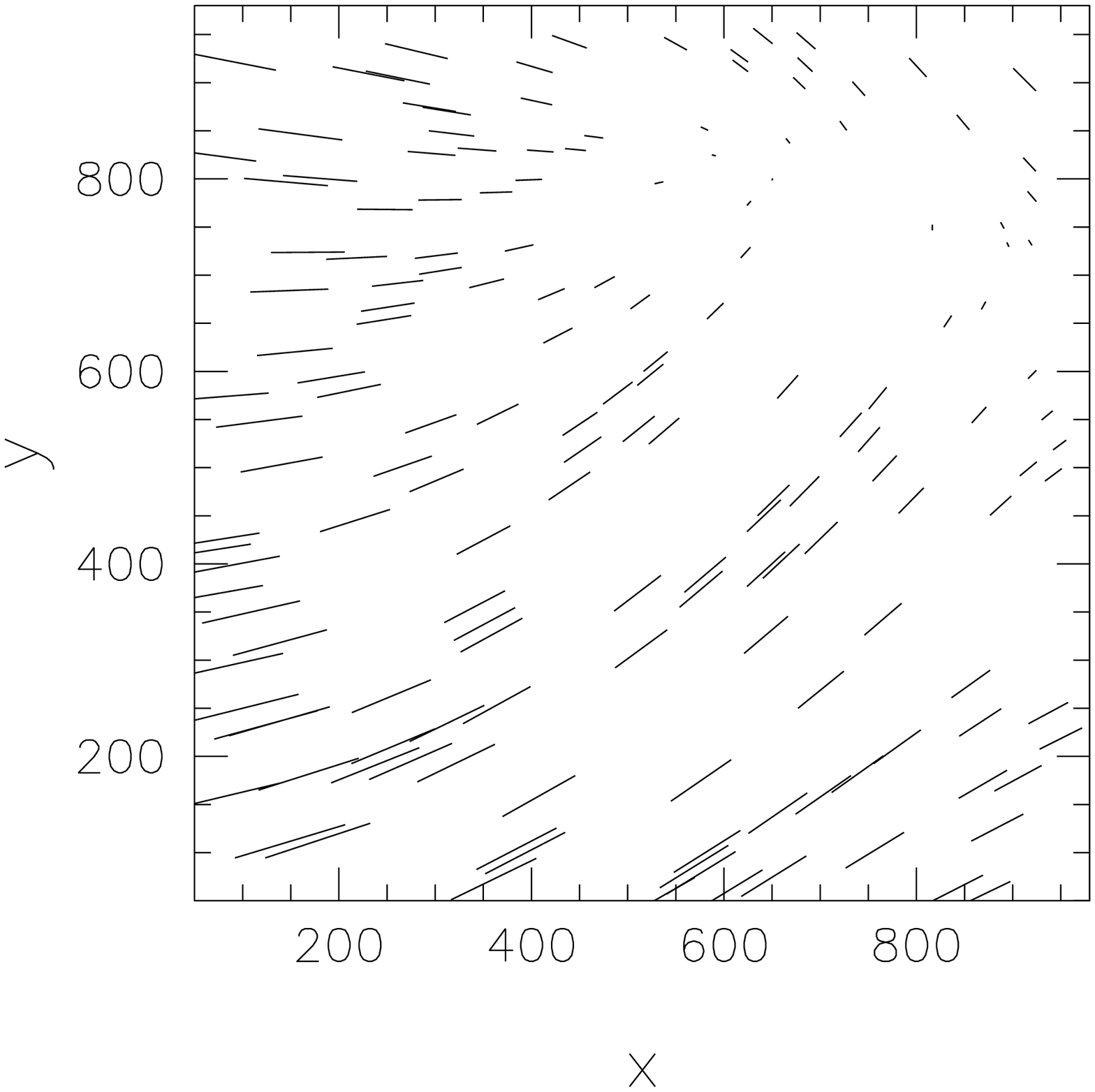}{4.6cm}
 \plotone{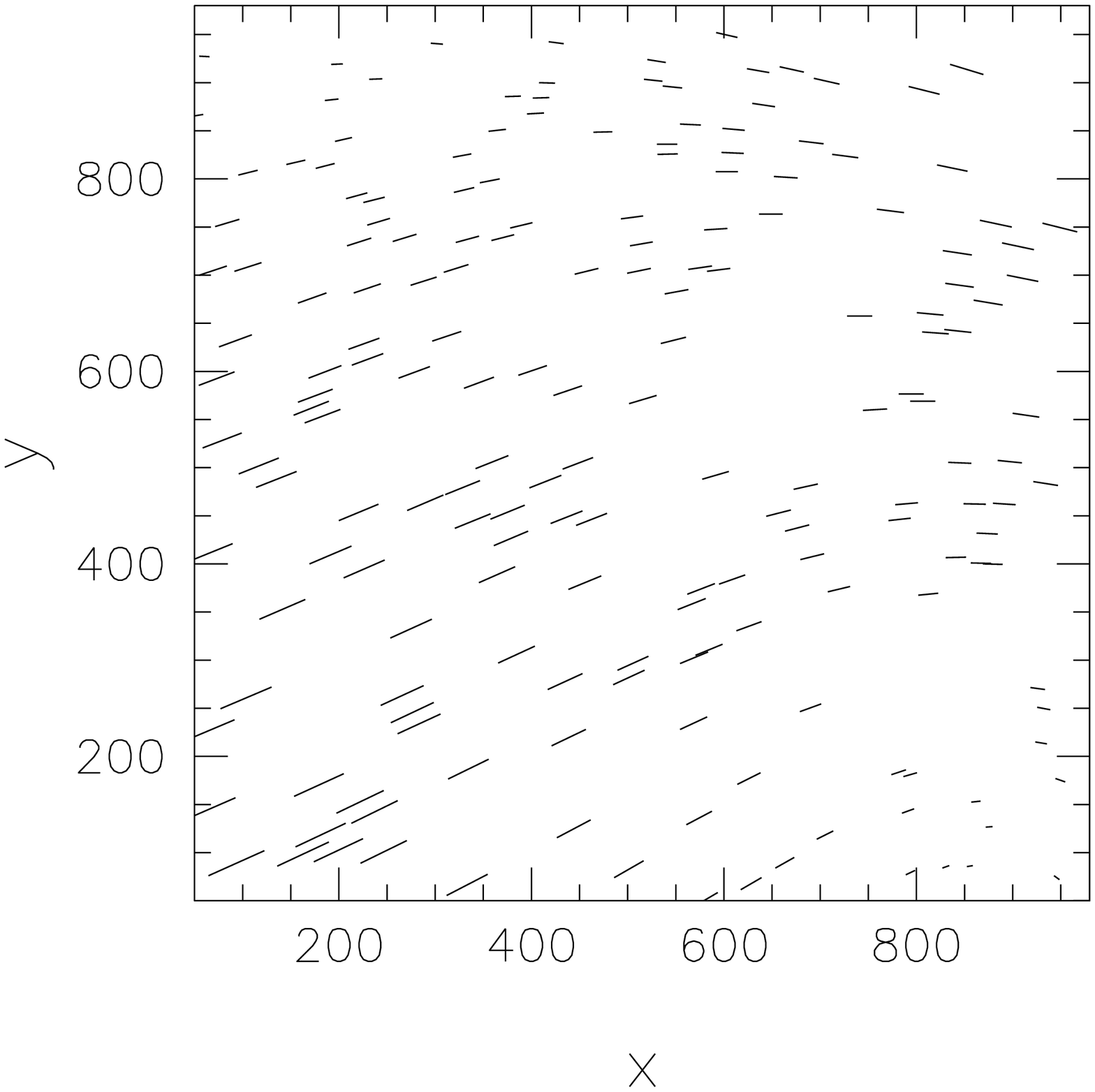}{4.6cm}
\end{minipage}\hspace{0.5cm}
\begin{minipage}{6cm}                      
\caption{For one star field the distribution of the ellipticities of
stars are shown before (top left) and after (top right) correcting
for PSF anisotropy. The middle left panel shows the spatial
distribution of the ellipticities across the STIS field, the middle
right panel the fitted second-order polynomial at the star
positions. The length of the sticks indicates the modulus of the
ellipticity, the orientation gives the position angle.  In the two
plots at the bottom we show 
the values of the polynomials for the anisotropy correction of two
individual, undrizzled exposures of the star field. The two individual
fields were 
taken with a time difference of only 30 minutes and demonstrate the
very short timescale variations of the PSF anisotropy pattern. }
\end{minipage}
\end{figure}

For the individual undrizzled images of some of the star fields we
also analysed the PSF anisotropy and find very short timescale
variations of the anisotropy pattern over the fields, as can be seen
in Fig.~2, bottom panels.

In Fig.~3 we plot the mean ellipticity for all galaxy fields 
with and without the PSF anisotropy correction. It illustrates
that the PSF anisotropy correction changes the mean ellipticity by an
amount typically smaller than $1\%$. Also, the dispersion between
different PSF models from different star fields is much less than
$1\%$, which means that the changes of the PSF anisotropy seen in
different star fields are sufficiently small to allow us to use one (or
a suitable combination of them) for the actual analysis.

The smearing corrected ellipticity of each galaxy is calculated by 
\begin{equation}
 e^{\mathrm{iso}} = (\Pg)^{-1} (e - P_{\mathrm{sm}} q^*),
\end{equation}
which is an unbiased (provided that $\ave{e_{\mathrm{s}}}=0$) but very
noisy estimate of the shear $\gamma$. 

To check that we did not introduce systematics in our galaxy selection
and PSF correction we also calculated the probability distribution for all
galaxy ellipticities which were used in the  cosmic shear analysis and
found that the mean ellipticity  is compatible with zero, and the
dispersion in both components is $\sigma_1=\sigma_2=26\%$. 

\begin{figure}
\begin{minipage}{9cm}
\plotone{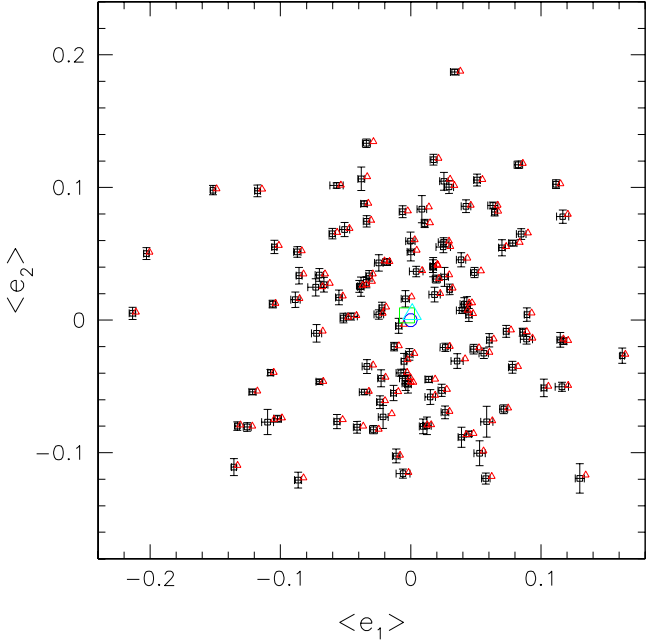}{9cm}
\end{minipage}\hspace{0.5cm}
\begin{minipage}{6.5cm}
\caption{For the galaxy fields we plot the mean uncorrected ellipticity
of galaxies (triangles) as well as the mean anisotropy corrected
ellipticity (squares). The error bars attached to the squares denote 3 times
the dispersion of the field-averaged corrected ellipticities when the
different PSF model fits are used. The error on the mean is much
smaller than the symbols used. The shift of the corrected mean
ellipticities toward negative $e_1$ is expected from the behaviour of
the stellar ellipticities plotted in Fig.~1. The big triangle and big
square in the centre denote the mean over all galaxy fields of the
uncorrected and corrected mean ellipticities, respectively; the size
of the symbols represent the $1\sigma$ errors on the mean. The circle
shows the origin for reference. }
\end{minipage}
\end{figure}

\section{Cosmic Shear}
The rms shear in a circular aperture  with angular radius $\theta$ is
related to the power spectrum of the surface mass density $\kappa$ by
\begin{equation}
 \cs (\theta) = 2\pi \int_0^\infty {\mathrm{d}}s \, s \, P_{\kappa}(s) \left[
 I(s\theta) \right]^2, 
\end{equation}
where $I(\eta):=\mathrm{J}_1(\eta)/(\pi \eta)$ and $\mathrm{J}_1$ is
the Bessel function of the first kind. $P_{\kappa}$ in turn is related
to the three-dimensional power spectrum by a simple projection:
\begin{equation}
P_{\kappa}(s) = \int_0^{w_{\mathrm{H}}} {\mathrm{d}}w \, g(w)
P\left(\frac{s}{f_{K}(w)}; w\right), 
\end{equation}
where $w$ is the  comoving distance,  $f_K(w)$ is the  comoving angular
diameter distance, $g(w)$ includes geometrical factors which depend on
the redshift distribution of the sources (see \cite{Sea98}), and
the integral extends to the horizon $w_{\mathrm{H}}$. 

Let $e_{in}$ denote the fully corrected ellipticity of the $i$-th
galaxy on the $n$-th field, then the quantity we measure for each field is 
\begin{equation}
 \csn := \frac{ \sum_{i\neq j} w_{in} w_{jn} e_{in}
 e_{jn}^\star} {\sum_{i\neq j} w_{in} w_{jn}} ,
\end{equation}
where $w_{in}$ is the weight of the $i$-th galaxy in the $n$-th
field. This is an unbiased estimate of the cosmic shear disperion in
the $n$-th field. Note that $\csn$ is not positive definite. From this
one obtains an unbiased estimate of the cosmic shear dispersion:
\begin{equation}
 \cs = \frac{\sum N_n \csn}{\sum N_n} ,
\end{equation}
where we weight each galaxy field by the number of galaxies per field
$N_n$ to minimize Poisson noise. 

Using all 121 galaxy fields we find an rms cosmic shear of $\sim 4\%$
with $1.5\sigma$ significance. 
Restricting the analysis to fields with higher number density of
galaxies we find a 
larger cosmic shear signal. This would agree
with a cosmological interpretation of the signal: Fields with a higher
number density of objects typically have a larger exposure time,
therefore they typically probe higher redshifts and one expects a
higher cosmic shear signal. 

If we apply the PSF corrections from different star fields
individually we find that the difference between the star field
corrections is much smaller than the statistical error on the cosmic
shear measurement, which again demonstrates that PSF effects of STIS
are small. 

In Fig.~4 we compare our result for the cosmic shear to
the ones obtained by other groups on larger scales and to the
theoretically expected values when using different cosmological models
with a mean source redshift of $\ave{z_{\mathrm{s}}}=1.2$, which is
appropriate for the ground-based measurements. With the STIS data we
are probably probing at higher mean redshifts, but since our
observations with STIS were taken in the CLEAR filter mode, we  do
not have any information about the exact redshift range of our
galaxies. Moreover, our fields have a large spread in exposure times
and therefore we effectively average over different cosmic shear
values. The galaxies on fields with longer exposure times are expected
to probe higher redshifts on average. Their light bundles traverse a larger
amount of matter, and therefore we expect a higher value for the
cosmic shear. Multicolour observations from the ground have been
carried out to determine the redshift distribution of galaxies in the STIS
fields photometrically. With these, we will be able to quantitatively
interpret the cosmic shear measurement relative to other angular scales. 

\pagebreak
\mbox{}
\vspace{-0.9cm}

Although we obtained a detection of the cosmic shear, the error bar
is still large. The error in the data depends both on the number of
galaxies per field and on the number of fields. It is therefore
important to get more fields with higher number densities of
objects. The parallel observations with STIS are currently continued
with a GO cycle 9 parallel proposal (8562+9248, PI: P.~Schneider).

\begin{figure}
\begin{minipage}{8cm}
\plotone{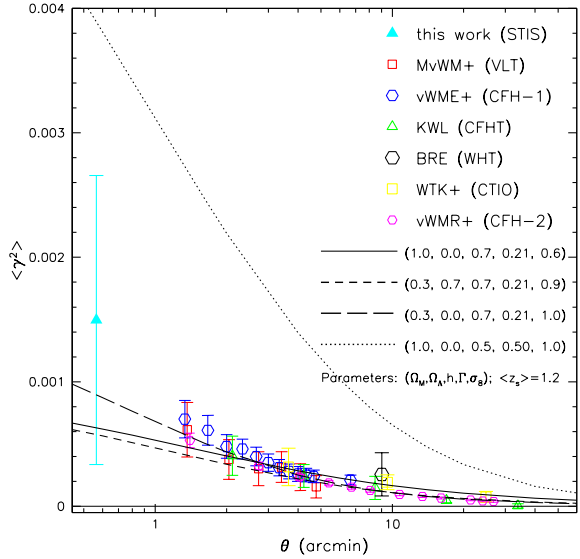}{8cm}
\end{minipage}\hspace{0.5cm}
\begin{minipage}{8cm}
\caption{Comparison of our cosmic shear result with measurements at
larger angular scales from other groups and with model
predictions. The lines show the theoretical predictions if one uses 
 different cosmological models, which are characterized by $\Omega_m$,
 $\Omega_\Lambda$, $h$, $\Gamma$ and $\sigma_8$. The redshift
 distribution is taken from \cite{BBS}, with a mean source redshift
 of $\ave{z_{\mathrm{s}}}=1.2$. }
\end{minipage}
\end{figure}

\mbox{}\\
\vspace{-1.35cm}

\acknowledgements{
We thank Y. Mellier and L. van Waerbeke for fruitful discussions.
This work was supported by the TMR Network ``Gravitational Lensing: New
Constraints on Cosmology and the Distribution of Dark Matter'' of the
EC under contract No. ERBFMRX-CT97-0172  and by the German
Ministry for Science and Education (BMBF) through the DLR under the
project 50 OR106.
}

\end{document}